\documentclass[12pt]{iopart}
\usepackage{iopams}  
\usepackage{graphicx}
\usepackage{color}  
\usepackage{cite}
\def\eqnref#1{Eq.~(\ref{#1})}
\def\figref#1{Fig.~(\ref{#1})}
\begin{document}

\title{A thermodynamic model for the O\(_2^-\) mobility in Ne gas in broad density and temperature ranges}

\author{A. F. Borghesani}
\address{CNISM Unit, Department of Physics and Astronomy\\ University of Padua, Italy}
\ead{armandofrancesco.borghesani@unipd.it}
\author{F.Aitken}
\address{University Grenoble Alpes, C.N.R.S., G2ELab, Grenoble, France}

\begin{abstract}
 We report new measurements of the mobility \(\mu\) of O$_{2}^{-}$ ions in supercritical neon  in the range $45\,\mathrm{K}\le T\le 334\,\mathrm{K}$ for   number density $N\ge 0.5\,$nm$^{-3}.$ 
 We rationalize the experimental data of all isotherms with the Stokes-Cunningham
formula by computing the ion hydrodynamic radius as a function of \(T \) and \(N\) with the thermodynamic free volume model  
 developed for the ion mobility in superfluid He.  
 The model parameters  are determined by re-analyzing published data for $T=45\,$K for \(N\) up to $N\approx 1.65N_{c} $ ($N_{c}\approx 14.4\,$nm$^{-3}$ is the critical number density), which roughly span four orders of magnitude of the Knudsen number $(0.1\le \mathrm{K}_{n}\le 1000)$, covering the transition from the kinetic- to the hydrodynamic transport regime. These parameters provide an excellent description of the dependence of 
\(\mu\) on \(N\) for all higher isotherms and 
yield a strict  
test of the model validity, thereby bridging the gap between the kinetic- and the hydrodynamic transport regimes.  
\end{abstract}
\vspace{2pc}
\noindent{\it Keywords}: negative ion mobility, supercritical Ne gas, free volume model. 
 
\maketitle

The knowledge of how ions drift through a gas or liquid under the action of an electric field 
is very important for both applications and fundamental science in physics and chemistry. The theory of ion transport is basic for many applications of low temperature plasmas~\cite{Bruggeman2016,Adamovich2017a,schmidt2005b} in biology, chemical synthesis, electrical discharges, high-energy physics~\cite{lopez2005}, and is also 
fundamental to shed light on the kinetic processes related to the complex interaction between charges and environment.
 In low-density gases the Kinetic Theory describes the ion mobility \(\mu\) via the momentum transfer scattering cross section determined by the ion-atom interaction potential~\cite{mason1988}. In high-density environments, the motion of thermal ions occurs in the hydrodynamic regime and the Stokes formula can be used to describe \(\mu\) by introducing an effective hydrodynamic radius \(R\). In this case the ions are used to probe the microscopic structure and behavior of liquids such as superfluid He (for a review, see ~\cite{Borghesani2007}), hydrocarbons~\cite{Gee1980,Huang1980}, and cryogenic liquids~\cite{Davis1962,Gee1985,Schmidt2005}. However, a unified theory (or model) describing how the ion transport behavior crosses over from the kinetic- to the hydrodynamic regime is still missing. Supercritical gases, whose density can be varied in an extremely broad range, offer a unique test field
to seek for a solution that bridges the gap between dilute gas and liquid.

Whereas the behavior of cations is amply investigated because are easily produced by direct ionization of the sample by means of high-energy radiation or 
electrical discharges, negative ions are not as thoroughly studied in spite of their relevance, e.g., in the physical chemistry of the atmosphere~\cite{mason2001}, because are produced by the quite inefficient, resonant, low-energy electron attachment process to electronegative molecular impurities~\cite{christophourou1984a}. A temporary anion formed in a vibrationally excited state~\cite{illenberger1994,Matejcik1996} rapidly decays by autodetachment unless it is stabilised by collisions with host gas atoms~\cite{Bloch1935,Bartels1973,Bruschi1984,borghesani1991,Neri1997a}. The processes leading to the 
formation of stable anions depend on the environment and give origin to states which cannot be adiabatically result by the addition of the ion to the environment~\cite{hernandez1991a} and 
whose structure is more complicated than that of cations~\cite{khrapak1999a}. Among electronegative impurities such as CO\(_{2}\)~\cite{nishikawa1999b}, NO~\cite{nishikawa1999}, and SF\(_{6}\)~\cite{Fehsenfeld1970}, O\(_{2}\)~\cite{Bradbury1933}, whose electron affinity is \(\approx 0.45\,\)eV~\cite{Travers1989}, plays the most important role because its ubiquitous presence as impurity even in the best purified gas makes the same ionic species, O\(_{2}^{-}\), available to probe sample specific features. The mobility of O\(_{2}^{-}\) has been investigated only in few experiments: in liquid Ar and Kr~\cite{Davis1962} and  Xe~\cite{Davis1962,Hilt1994,Schmidt2005} and in supercritical He\cite{Bartels1975,Borghesani1995}, Ne~\cite{Borghesani1993,Borghesani1995}, and Ar~\cite{Borghesani1997,Borghesani1999}. Here, we report new results obtained  in Neon gas on several isotherms in a broad density range and show how the experimental results are very well rationalized 
by a recent thermodynamical model aimed at predicting the ion hydrodynamic radius\cite{Aitken2011,Aitken2011a,Aitken2015,Tarchouna2015,Aitken2017}.

\begin{figure}[t!]\centering\includegraphics[width=\textwidth]{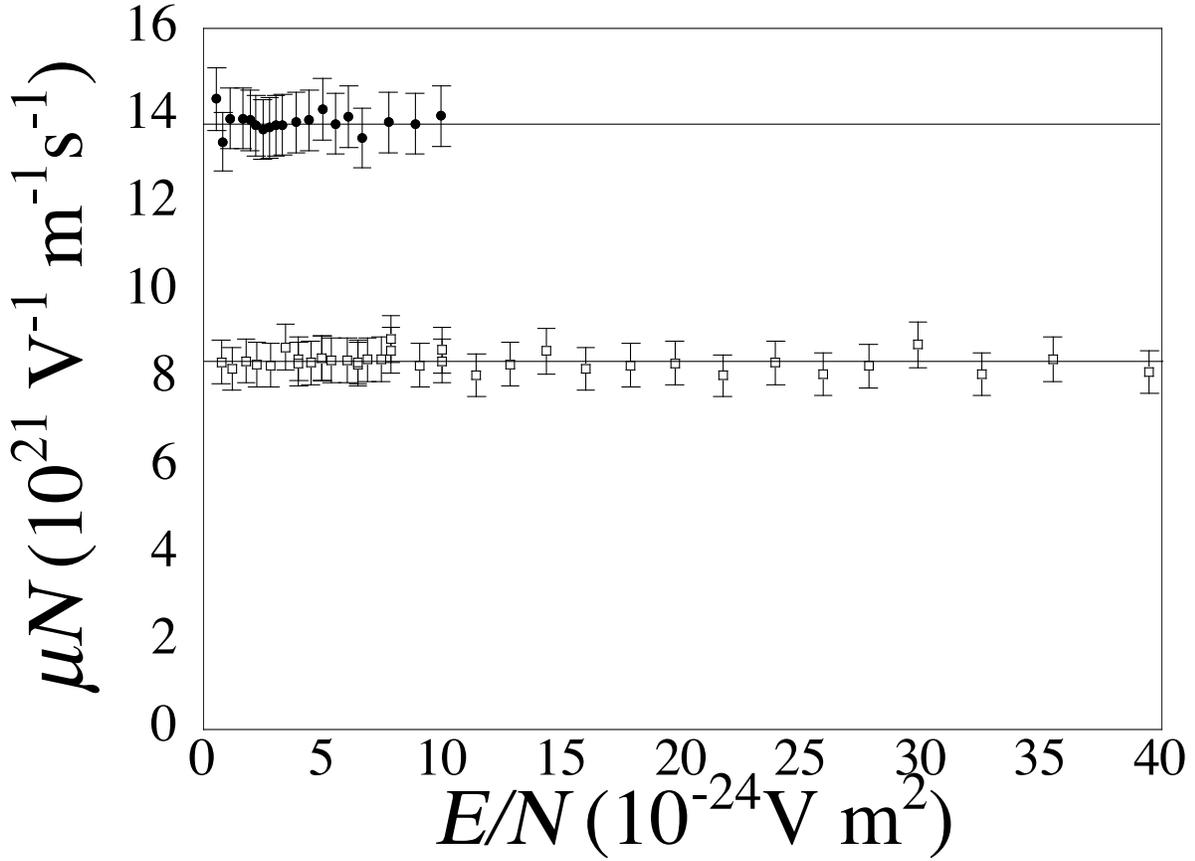} \caption{\small 
\(\mu N\) vs \(E/N\) of  
O\(_{2}^{-}\) in Ne gas at \(T=45\,\)K. Closed points: \(N=23.45\,\)nm\(^{-3}\). Open points: \(N=2.63\,\)nm\(^{-3}\).\label{fig:figure1}}
\end{figure}

The experiment is based on the pulsed Townsend electron photoinjection technique. The technical details of the experiments are described in literature~\cite{Borghesani1993}. 
Ions are produced by resonant electron attachment to O\(_{2}\) impurities in a concentration of a few tens of ppm. The drift field is so weak that the density normalized field \(E/N\) never exceeds a few tens of mTd \((1\,\mbox{Td}\,=10^{-21}\,\mbox{V}\,\mbox{m}^{2})\). Thus, the ions always are in thermal equilibrium with the gas 
and \(\mu\) does not depend on \(E/N\), as shown in \figref{fig:figure1}.

\begin{figure}[t!]
\centering
\includegraphics[width=\textwidth]
{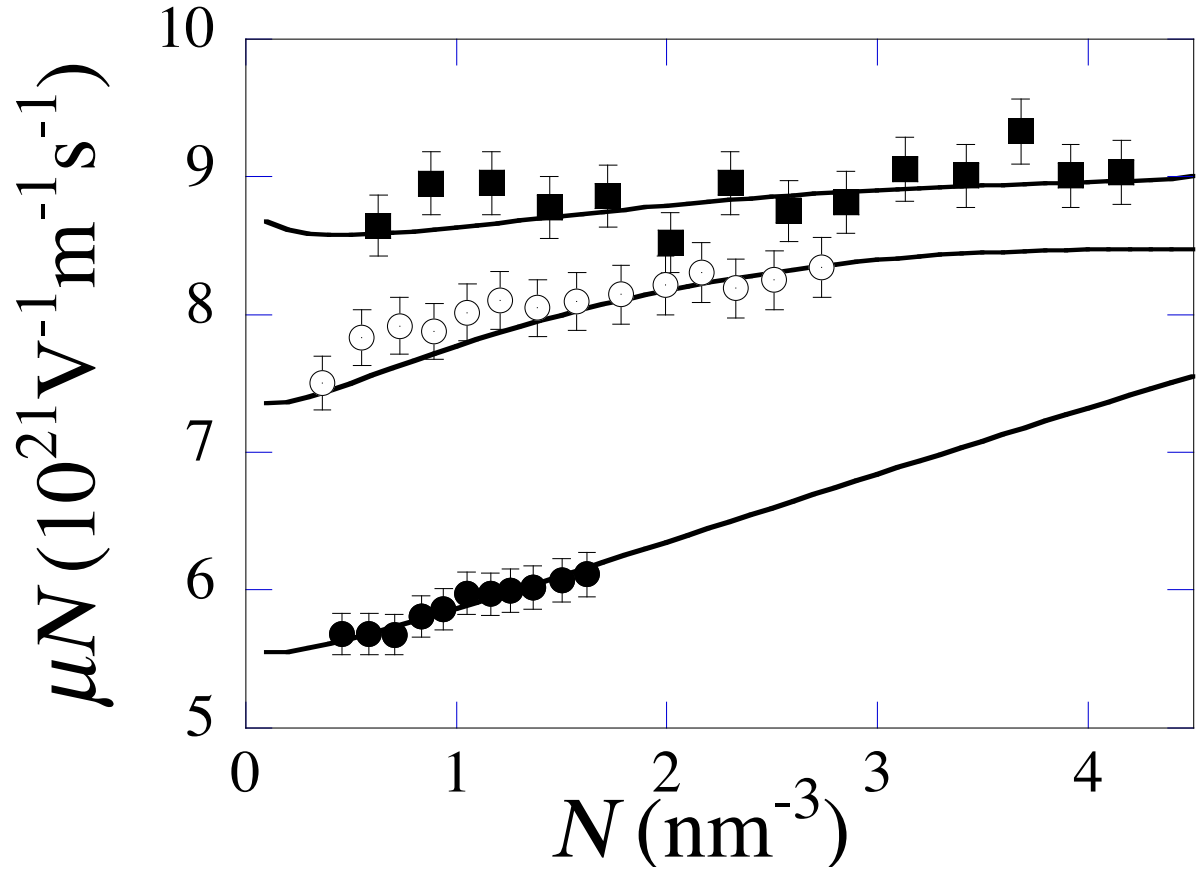}
\caption{\small \(\mu N\) vs \(N\) for some isotherms.
From top: \(T=130.8\,\), \(64.4\,\), and \(334\,\)K. Lines: model prediction.\label{fig:figure2}}
\end{figure}

The theoretical interpretation of such measurements is still unsatisfactory. 
In the gas at low- up to intermediate \(N\) along isotherms the experimental density normalized mobility \(\mu N\)  showed a small, though unmistakable, almost linear dependence on \(N\)~\cite{Borghesani1995}.  This dependence is also present in the  new measurements displayed in \figref{fig:figure2} 
whereas, according to the Kinetic Theory~\cite{Kihara1953}, \(\mu N\) should be constant
\begin{equation}
\mu N =\frac{3e}{8\pi R^{2} }\left(
\frac{\pi}{2m_{r}k_\mathrm{B}{T}}
\right)^{1/2}
\label{eq:muNkinetic}
\end{equation}
 in which \(R\) is the hard-sphere radius of the ion and \(m_{r}\) is the ion-atom reduced mass. Moreover, the low density limit of \(\mu N\) disagrees by a factor \(\approx 2\) with the Langevin's prediction~\cite{langevin1905,mason1988,Borghesani1993}. 

\begin{figure}[b!]
    \centering
    \includegraphics[width=\textwidth]{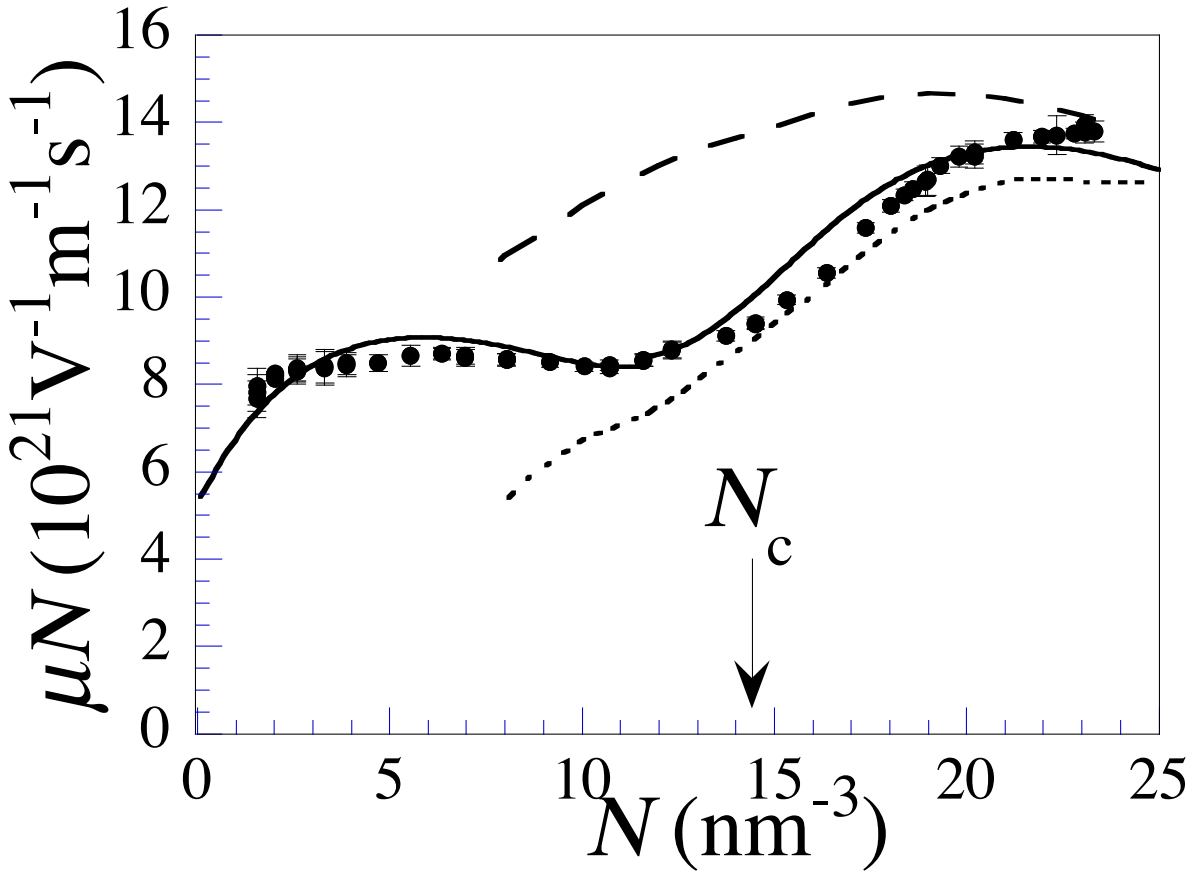}
    \caption{\small \(\mu N\) vs \(N\) for O\(_{2}^{-}\) ions in Ne gas at \(T=45\,\)K~\cite{Borghesani1993}. Dashed line: solution of the NS equation accounting for the electrostriction induced nonuniformities. Dotted line: Khrapak's model~\cite{Volykhin1995}. Solid line: present model.\label{fig:figure3}}
\end{figure}

By contrast, at densities comparable to those of a liquid, the ion mobility cannot simply be described by the hydrodynamic Stokes formula with a constant hydrodynamic radius \(R\)
\begin{equation}
\mu = \frac{e}{6\pi\eta R}
\label{eq:mustokes}\end{equation} in which \(\eta\) is the 
viscosity, even if the Navier-Stokes (NS) equations are solved by taking into account the density and viscosity non-uniformities induced by electrostriction 
~\cite{Ostermeier1972,Borghesani1993} (dashed line in \figref{fig:figure3}, in which the experimental data for \(T=45\,\)K~\cite{Borghesani1993} are shown).

A great improvement of the rationalization of the experiments has been achieved by accounting for the structure of the O\(_{2}^{-}\) ion~\cite{Khrapak1995,Volykhin1995}. 
Besides the solvation shell formed by electrostriction~\cite{Atkins1959}, the additional electron interacts with the electronic shells of the surrounding atoms via repulsive exchange forces that overwhelm the attractive polarization interaction and a void is created around the bare ion. So, the transport is determined by the interaction of such large complex structure with the gas and is quite insensitive to the ionic species. Detailed calculations~\cite{Khrapak1995,Volykhin1995} yield the void radius \(R\) by minimizing the free energy of the complex and obtain 
 the gas density profile established around the ion. 
 \(R\) turns out to be almost independent of  
 \(N\) whereas the local density profile induced by electrostriction strongly depends on the 
 polarizability of the gas. 
 In Ne the measurements were carried out \(T\approx 45\,\)K~\cite{Borghesani1993}, i.e., very close to the critical temperature when the gas is most compressible. As a consequence of its large  
 polarizability, size and density of the solvation shell cannot be neglected. At the largest \(N\) the mean free path and the mean interatomic separation are much smaller than the cavity size and the hydrodynamic Stokes formula~\eqnref{eq:mustokes} is used to describe \(\mu\). 
 The position of the maximum of the density profile is chosen as hydrodynamic radius 
 and \(\eta\) is evaluated at the density of the maximum. The agreement with the data is quite reasonable for \(15\,\)nm\(^{-3}\le N\le 25\,\)nm\(^{-3}\)  but it fails at reproducing the data for smaller \(N\) and the crossover between Knudsen- and hydrodynamic regime~\cite{Volykhin1995} (dotted line in \figref{fig:figure3}). 

Recently, a heuristic model has appeared which successfully rationalizes electron and cation mobility in liquid and supercritical He~\cite{Aitken2011,Aitken2011a,Aitken2015,Tarchouna2015,Aitken2017}. 
The model is based on the  free volume concept 
adopted in the past to 
predict properties such as the ionic conductivity in polymers~\cite{Miyamoto1973} and  the viscosity of liquids~\cite{Doolittle1951}. The model provides a thermodynamic equation of state (EOS) 
for the \(N\) and 
\(T\) dependence of the free volume \(V_{f}\) available for the ion motion. The linear size of the free volume per particle is assumed to be related to the effective radius in the Stokes formula. Electrons in liquid helium are fully solvated and are surrounded by a hollow cavity so that the volume per solvated particle is \(V_{s}\propto V_{f}/N\), where \(V_{f}=V-b\). \(V\) is the macroscopic volume occupied by the medium and 
\( b\) is the total solvent covolume. The covolume \(b^{\prime}\) of ions is neglected because of their extremely low concentration. The thermodynamic description of \(V_{s}\) is accomplished by a van der Waals-like EOS
\begin{equation}
V_{s}= \frac{k_\mathrm{B}T}{P+\Pi}
\label{eq:EOSvdW}\end{equation}
in which \(P\) is the ordinary pressure and \(\Pi\) is the internal pressure that accounts for the attractive potential energy contributions in the system. 
For a fully developed solvation shell of spherical shape,  its linear size is simply given by \(R - R_{0}=\left({3 V_{s}}/{4\pi}\right)^{1/3}\), where  \(R_{0}\) is the hard-sphere radius of the charge-medium interaction. 
By contrast, for ions in a supercritical gas, compressibility effects have to be taken into account. At constant \(T\)
, the solvation shell size follows the 
behavior of the compressibility~\cite{Itoh2001,Itoh2004}.
Starting from low \(N\) 
the 
shell is gradually growing up along with the compressibility. At higher \(N\)  
compressibility and size reach a maximum and then decrease with a further density increase whereas the free volume is a monotonic function of \(N\). Thus, we expect that both 
\(V_{s}\) and \( R-R_0 = g(V_{f}/N)\) are functions of \(V_{f}/N\). The analytical forms of \(\Pi\) and \(g\) have to be determined by enforcing agreement with the experimental results.

At high \(N\), where the Knudsen number, , i.e., the ratio of the atom mean free path \(\ell\) to the size of the ion \(R\), is \(K_{n}=\ell/R\ll 1\),  the ion drift occurs in the hydrodynamic regime described by the Stokes formula. However, as most of the experimental data fall in the transition region from the hydrodynamic- to the kinetic regime in which \(K_{n}\gg 1\), 
we extend the predictive range of the model by using the empirical Millikan-Cunningham interpolation formula for the mobility~\cite{Cunningham1910,Millikan1910,Tyndall1938}
\begin{equation}
\mu N = \frac{eN}{6\pi \eta R}\left\{1+ \phi\left[K_{n}(T,N)\right]\right\}
\label{eq:muninterp}\end{equation}
This approach has quite successfully been pursued, for instance, for localized electrons in dense 
He~\cite{Levine1967} and Ne gas~\cite{Borghesani1990}.
The slip correction factor \(\phi\) must vanish at high \(N\) and, for large \(K_{n}\) at low \(N\), it must be such that \(\mu\) is given by \eqnref{eq:muNkinetic}. Several analytic forms are proposed in literature (for a review, see~\cite{Allen1982}), but we have chosen the present one by enforcing agreement with the data.
 
The published Neon data at \(T=45\,\)K~\cite{Borghesani1993} span a broad density range \(0.17 \lesssim N/N_{c}\lesssim 1.7\) that almost entirely covers the crossover region and their re-analysis 
 allows us to fix all the model parameters.  According to the Occam's razor principle~\cite{Priest1981} we have tried to keep their number as small as possible. 
Firstly, the simplest choice for the internal pressure term far away from any phase transition is 
\begin{equation}
\Pi (N,T) = \alpha N^{2}
\label{eq:internal}\end{equation}
The \(\alpha\) parameter determines the density of the compressibility maximum.
The experimental data show a shallow minimum around \(N_{m}\approx 11.5\,\)nm\(^{-3}.\) Thus, it is reasonable to assume that the effective ion radius is maximum for this density value where the system compressibility is the largest. This observation yields \(\alpha =8.937\times 10^{5}\,\)Pa\(\,\)nm\(^{6}\). It is extremely interesting to note that the same value gives good agreement with the data in liquid and supercritical He~\cite{Aitken2011,Aitken2011a,Aitken2015,Tarchouna2015,Aitken2017}. 
 \(\alpha\), which is related to the attractive interactions, 
 appears to be system independent, being thus endowed with some sort of universal character.
 
 We expect that the effective ion radius at constant $T$ is given by a functional form of the scaled free volume \(V_{0}/V_{s}\), in which \(V_{0}\) belongs to a suitable reference state, identical to that which allowed the rationalization of other mobility measurements\cite{Aitken2011,Aitken2011a,Aitken2015,Tarchouna2015,Aitken2017}
\begin{equation}
\frac{R}{R_{0}} = 1 +   \frac{\left(V_{0}/V_{s}\right)^{\epsilon_{1}}}{1+\gamma \left(V_{0}/V_{s}\right)^{\epsilon_{1}}\exp{\left[-K(T)\left(V_{0}/V_{s}\right)^{\epsilon_{2}}\right]} }
\label{eq:rover0}\end{equation}
By casting \eqnref{eq:EOSvdW} in the form
\begin{equation}
\frac{V_{0}}{V_{s}}= \delta \frac{P+\Pi}{T}
\label{eq:v0suvs}\end{equation} 
with \(\delta=0.15\), a fit of the mobility data at the highest \(N\) gives the values \(R_{0}\approx  0.315\,\)nm, \(\epsilon_{1}=1/2,\) \(\epsilon_{2}=-1\), and \(\gamma = 6.3\). For the Neon EOS  
we used literature data~\cite{katti}. We 
emphasize that the analytical form of \eqnref{eq:rover0} has been determined by investigating several possibilities and picking the simplest form provided that it minimizes the deviations from the data. We also note that this analytical form is valid for all other mobility measurements~\cite{Aitken2011,Aitken2011a,Aitken2015,Tarchouna2015,Aitken2017}. 

 An analysis of the data for \(T>45\,\)K shows that \(K(T)\) roughly describes an exponentially decreasing function of \(T/T_{c}\). Its best form is
\begin{equation}
K(T)=k_{1}\mathrm{e}^{-\zeta T/T_{c}}+k_{2}\mathrm{e}^{-\left(\vert T/T_{c}-\xi\vert/2\right)^{3}}
\label{eq:kot}\end{equation}
with \(k_{1}=148.05\), \(k_{2}=2.0285\), \(\zeta = 2.0279\), and \(\xi=3.\)
 Its physical meaning is that it makes the effective radius more sensitive to the free volume at low \(T\) whereas it guarantees that, at higher \(T\) and low \(N\),  the effective radius roughly tends to the hard-sphere radius of the ion-atom interaction potential.

For the slip factor correction we used a modified version of the formula proposed by Knudsen {\em et al.}~\cite{Knudsen1911,Kim2005} in order to account for objects whose size may depend on \(T\) and \(N\)
\begin{equation}
\phi (N,T)= \frac{1}{2}\frac{N_{c}}{N}\mathrm{e}^{-\left[ \chi(T)/ K_{n}(N,T)\right]^{3/4}} \mathrm{e}^{T_{c}/T}\left(1-\mathrm{e}^{-T/T_{c}}\right)
\label{eq:phi}
\end{equation}
In order to evaluate the Knudsen number we used the experimental determination of the mean free path from the viscosity as prescribed by the Kinetic Theory~\cite{Reif}
\begin{equation}
\ell=\frac{3\eta(N,T)}{Nm_{r}\bar v}
\label{eq:kappaenne}\end{equation}
in which
\(\bar v= \left[8k_\mathrm{B}T/\pi m_{r}\right]^{1/2}\) is the mean thermal velocity. \(\eta\) is computed with NIST software.

The analysis of the data at higher \(T\) requires the introduction of a temperature dependence in the slip factor by the function \(\chi (T)\) which is approximately constant in the \(65\,\mbox{K}\,<T<250\,\mbox{K}\) range  but slightly increases the correction factor at higher \(T\) and reduces it at lower \(T\). Its role can heuristically be  understood by noting that it favors the ion hydrodynamic drift behavior at lower \(T\) where the formation of a solvation shell is easier.
Its best form is
\begin{equation}
\chi(T) =\chi_{1}\mathrm{e}^{-\zeta_{2}T/T_{c}}+\chi_{2}\mathrm{e}^{-\left(\vert T/T_{c}-\xi_{2}\vert/ \zeta_{2}
\right)^{\chi_{3}}}
\label{eq:chiT}\end{equation}
with \(\chi_{1}=36.479,\) \(\zeta_{2}= 2.6931,\) \(\chi_{2}= 0.957,\) \(\xi_{2}= 3.2117\), and \( \chi_{3}=7.8475.\)
 At the same time the factor \(\mathrm{e}^{T_{c}/T}(1-\mathrm{e}^{-T/T_{c}})\)
has been included in order to obtain agreement with the temperature dependent, low-density limit of the experimental data.

The model prediction for \(T=45\,\)K is the
solid line in \figref{fig:figure3}. It is interesting to note that the agreement with the data, though still not perfect, is now obtained over the entire density range thereby bridging the kinetic- and hydrodynamic transport regimes.

\begin{figure}[b!]\centering\includegraphics[width=\textwidth]{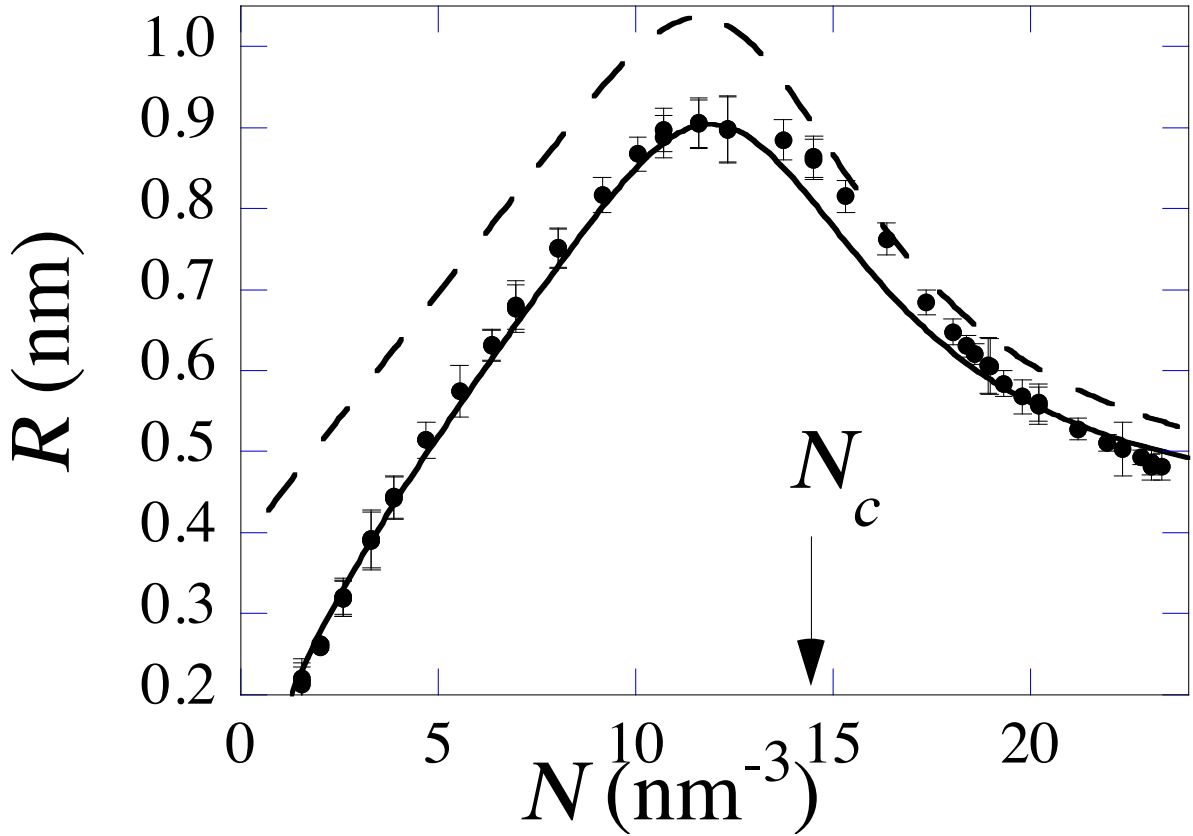}
\caption{\small \(R\) 
vs \(N\) for $T=45\,$K. Dashed line: \eqnref{eq:rover0}. Solid line: \eqnref{eq:reff}. Points: experimental determination. 
\label{fig:figure4}}
\end{figure}

It also interesting to compare in \figref{fig:figure4} the hydrodynamic radius given by \eqnref{eq:rover0} and the effective hydrodynamic radius 
\begin{equation}
R_\mathrm{eff}= \frac{R}{1+\phi(N,T)}
\label{eq:reff}\end{equation} with its experimental determination 
obtained by inverting the Stokes formula. 
 First of all, we note that the experimental radius shows a maximum at precisely the same density of the shallow minimum in \(\mu N\). The hydrodynamic radius computed by \eqnref{eq:rover0} shares the same behavior although it is more than 20\(\,\)\% larger.  
 However, if the slip factor is accounted for, the experimental and the computed radii are in very good agreement. 
The position of the maximum is fixed by the choice of the
 \(\alpha\) parameter 
 that contains all the contributions of the attractive potential energy in the system.  The compressibility is 
maximum when the attraction and repulsion forces between atoms are in balance so that the atoms can more easily rearrange themselves to build local structures without too large a free energy cost~\cite{Itoh2004}. 
  Actually, the maximum compressibility of the pure gas occurs for \(N_c\), whereas the presence of the strong attractive ion-atom polarization 
interaction  
increases the gas compressibility by shifting its maximum to a lower \(N\).

Although all parameters in the free volume model have 
 mainly been determined by seeking for agreement with the low \(T\) data, we have applied the model to all 15 investigated isotherms by keeping the parameters fixed at the values determined at low \(T.\) In \figref{fig:figure2}
we compare the  
model prediction with the experimental data on three higher isotherms.
We see that the 
model is remarkably successful at describing the \(N\) and \(T\)  dependence of the experimental data.

In conclusion, we have measured the 
mobility of Oxygen ions in supercritical Neon gas for \(45\,\mbox{K}<T<334\,\)K and 
adopted the free volume model to compute the hydrodynamic radius of the complex that is formed as a balance between long range attractive polarization interaction and short range repulsive exchange forces when an ion is embedded in a fluid. 
The variation of the hydrodynamic radius that at first increases with \(N\), reaches a maximum for 
\(N_{m}< N_{c}\),  
and then decreases as \(N\) 
is increased further can be interpreted  as the condensation, growth and compression of negatively charged clusters~\cite{Tarchouna2015}.  Remarkably, in the present case the clusters are formed around a cavity surrounded by a solvation shell in which a negative ion is embedded.  Thus, the measurements of the O$_2^-$ mobility in a non polar solvent yield pieces of information specific to the solvation shell because the ion core can safely be expected to stay unchanged owing to the strong short-range repulsive exchange forces and any change of the cluster size has to be attributed to changes of the thickness of the solvation shell itself.

In order that the description of the mobility bridges the low-density kinetic transport regime and the high-density hydrodynamic one we used a modified version of the Millikan-Cunningham slip factor. We have to stress the fact that
a good description of the density dependence of the mobility data on all isotherms is obtained by using the parameter values  
that are mainly determined from the analysis of the low temperature data. In spite of the fact that some experimental observations cannot fully be explained
by the present approach, such as a residual, extremely weak field dependence of \(\mu\) at low \(N\) and high \(T\), we are confident that these results are promising to extend the investigation to negative ions in other noble gases as well as to positive ions in different 
fluids.


\section*{Acknowledgments}
The authors wish to thank K. von Haeften for critical reading of the manuscript.
\section*{References}
\providecommand{\newblock}{}


\begin{thebibliography}{10}
\expandafter\ifx\csname url\endcsname\relax
  \def\url#1{{\tt #1}}\fi
\expandafter\ifx\csname urlprefix\endcsname\relax\def\urlprefix{URL }\fi
\providecommand{\eprint}[2][]{\url{#2}}

\bibitem{Bruggeman2016}
Bruggeman P~J et al. 
 2016 {\em Plasma Sources Sci. Technol.\/} {\bf 25} 053002

\bibitem{Adamovich2017a}
Adamovich I et al. 
  2017 {\em J. Phys. D. Appl. Phys.\/} {\bf 50} 323001

\bibitem{schmidt2005b}
Schmidt W~F and Yoshino K 2005 {\em {Electronic Excitations in Liquefied Rare
  Gases}\/} (Stevenson Ranch, CA (USA): American Scientific Publishers) chap
  {Electric Discharges}, pp 296--315

\bibitem{lopez2005}
Lopez I~M and Chepel V 2005 {\em {Electronic Excitations in Liquefied Rare
  Gases}\/} (Stevenson Ranch, CA (USA): American Scientific Publishers) chap
  {Rare Gas Liquid Detectors}, pp 331--388

\bibitem{mason1988}
Mason E~A and McDaniel E~W 1988 {\em {Transport Properties of Ions in Gases}\/}
  (New York: Wiley)

\bibitem{Borghesani2007}
Borghesani A~F 2007 {\em {Ions and Electrons in Liquid Helium}\/} ({
  International Series of Monographs on Physics\/} vol 137) (Oxford: Oxford
  University Press)

\bibitem{Gee1980}
Gee N and Freeman G~R 1980 {\em Can. J. Chem.\/} {\bf 58} 1490--1494

\bibitem{Huang1980}
Huang S~S and Freeman G~R 1980 {\em J. Chem. Phys.\/} {\bf 72} 1989--1993

\bibitem{Davis1962}
Davis H~T, Rice S~a and Meyer L 1962 {\em J. Chem. Phys.\/} {\bf 37} 2470--2472

\bibitem{Gee1985}
Gee N, Floriano M~A, Wada T, Huang S~S~S and Freeman G~R 1985 {\em J. Appl.
  Phys.\/} {\bf 57} 1097--1101

\bibitem{Schmidt2005}
Schmidt W~F, Hilt O, Illenberger E and Khrapak A~G 2005 {\em Radiat. Phys.
  Chem.\/} {\bf 74} 152--159

\bibitem{mason2001}
Hughes P and Mason N 2001 {\em {Introduction to Environmental Physics: Planet
  Earth, Life and Climate}\/} (Boca Raton: CRC Press)

\bibitem{christophourou1984a}
Christophorou L~G, McCorkle D~L and Christodoulides A~A 1984 {\em
  {Electron-Molecule Interactions and Their Applications}\/} vol~I (Orlando:
  Academic Press) chap {Electron Attachment Processes}

\bibitem{illenberger1994}
Illenberger E 1994 {\em {Linking the Gaseous and Condesed Phases of Matter, The
  Behavior of Slow Electrons}\/} ({\em NATO ASI Series B: Physics\/} vol 326)
  (New York: Plenum Press) chap {Electron Attachment to Molecules}, pp 355--376

\bibitem{Matejcik1996}
Matejcik S, Kiendler A, Stampfli P, Stamatovic A and M{\H{a}}rk T~D 1996 {\em
  Phys. Rev. Lett.\/} {\bf 77} 3771--3774

\bibitem{Bloch1935}
Bloch F and Bradbury N~E 1935 {\em Phys. Rev.\/} {\bf 48} 689--695

\bibitem{Bartels1973}
Bartels A 1973 {\em Phys. Lett. A\/} {\bf 45} 491--492

\bibitem{Bruschi1984}
Bruschi L, Santini M and Torzo G 1984 {\em J. Phys. B At. Mol. Phys.\/} {\bf
  17} 1137--1154

\bibitem{borghesani1991}
Borghesani A~F and Santini M 1991 {Electron Localization Effects and Resonant
  Attachment to O\(_2\) Impurities in Highly Compressed Neon Gas} {\em Gaseous
  Dielectrics VI\/} ed Christophorou L~G and Sauers I (New York: Plenum Press)
  pp 27--33

\bibitem{Neri1997a}
Neri D, Borghesani A~F and Santini M 1997 {\em Phys. Rev. E\/} {\bf 56}
  2137--2142

\bibitem{hernandez1991a}
Hernandez J~P 1991 {\em Rev. Mod. Phys.\/} {\bf 63} 675--697

\bibitem{khrapak1999a}
Khrapak A~G, Tegeder P, Illenberger E and Schmidt W~F 1999 {\em Chem. Phys.
  Lett.\/} {\bf 310} 557--560

\bibitem{nishikawa1999b}
Nishikawa M, Itoh K and Holroyd R~A 1999 {\em J. Phys. Chem. A\/} {\bf 103}
  550--556

\bibitem{nishikawa1999}
Nishikawa M, Holroyd R and Itoh K 1998 {\em J. Phys. Chem. B\/} {\bf 102}
  4189--4192

\bibitem{Fehsenfeld1970}
Fehsenfeld F~C 1970 {\em J. Chem. Phys.\/} {\bf 53} 2000--2004

\bibitem{Bradbury1933}
Bradbury N~E 1933 {\em Phys. Rev.\/} {\bf 44} 883--890

\bibitem{Travers1989}
Travers A~J, Cowles D~C and Ellison G~B 1989 {\em Chem. Phys. Lett.\/} {\bf
  164} 449--455

\bibitem{Hilt1994}
Hilt O, Schmidt W~F and Khrapak A~G 1994 {\em IEEE Trans. Dielectr. Electr.
  Insul.\/} {\bf 1} 648--656

\bibitem{Bartels1975}
Bartels A 1975 {\em Appl. Phys.\/} {\bf 8} 59--64

\bibitem{Borghesani1995}
Borghesani A~F, Chiminello F, Neri D and Santini M 1995 {\em Int. J.
  Thermophys.\/} {\bf 16} 1235--1244

\bibitem{Borghesani1993}
Borghesani A~F, Neri D and Santini M 1993 {\em Phys. Rev. E\/} {\bf 48}
  1379--1389

\bibitem{Borghesani1997}
Borghesani A~F, Neri D and Barbarotto A 1997 {\em Chem. Phys. Lett.\/} {\bf
  267} 116--122

\bibitem{Borghesani1999}
Borghesani A, Neri D and Barbarotto A 1999 {\em Int. J. Thermophys.\/} {\bf 20}
  899--909

\bibitem{Aitken2011}
Aitken F, Li Z~L, Bonifaci N, Denat A and von Haeften K 2011 {\em Phys. Chem.
  Chem. Phys.\/} {\bf 13} 719--724

\bibitem{Aitken2011a}
Aitken F, Bonifaci N, Denat A and {Von Haeften} K 2011 {\em J. Low Temp.
  Phys.\/} {\bf 162} 702--709

\bibitem{Aitken2015}
Aitken F, Bonifaci N, Mendoza-Luna L~G and von Haeften K 2015 {\em Phys. Chem.
  Chem. Phys.\/} {\bf 17} 18535--18540

\bibitem{Tarchouna2015}
Tarchouna H~G, Bonifaci N, Aitken F, {Mendoza Luna} L~G and von Haeften K 2015
  {\em J. Phys. Chem. Lett.\/} {\bf 6} 3036--3040

\bibitem{Aitken2017}
Aitken F, Volino F, Mendoza-Luna L, Haeften K and Eloranta J 2017 {\em Phys.
  Chem. Chem. Phys.\/} {\bf 19} 15821--15832

\bibitem{Kihara1953}
Kihara T 1953 {\em Rev. Mod. Phys.\/} {\bf 25} 844--852

\bibitem{langevin1905}
Langevin M~P 1905 {\em Ann. Chim. Phys.\/}  245--288

\bibitem{Ostermeier1972}
Ostermeier R~M and Schwarz K~W 1972 {\em Phys. Rev. A\/} {\bf 5} 2510--2519

\bibitem{Volykhin1995}
Volykhin K~F and Khrapak A~G 1995 {\em JETP\/} {\bf 81} 901--908

\bibitem{Khrapak1995}
Khrapak A~G, Schmidt W~F and Volykhin K~F 1995 {\em Phys. Rev. E\/} {\bf 51}
  4804--4806

\bibitem{Atkins1959}
Atkins K~R 1959 {\em Phys. Rev.\/} {\bf 116} 1339--1343

\bibitem{Miyamoto1973}
Miyamoto T and Shibayama K 1973 {\em J. Appl. Phys.\/} {\bf 44} 

\bibitem{Doolittle1951}
Doolittle A~K 1951 {\em J. Appl. Phys.\/} {\bf 22} 1471--1475 

\bibitem{Itoh2001}
Itoh K, Holroyd R~a and Nishikawa M 2001 {\em J. Phys. Chem. A\/} {\bf 105}
  703--709

\bibitem{Itoh2004}
Itoh K, Muraoka K, Nagata T, Nishikawa M and Holroyd R 2004 {\em J. Phys. Chem.
  B\/} {\bf 108} 10177--10184

\bibitem{Cunningham1910}
Cunningham E 1910  {\bf 83} 357--365

\bibitem{Millikan1910}
Millikan R~A 1910 {\em Science\/} {\bf 32} 436--448

\bibitem{Tyndall1938}
Tyndall A~M 1938 {\em The mobility of positive ions in gases\/} (Cambridge: The
  University Press)

\bibitem{Levine1967}
Levine J~L and Sanders T~M~J 1967 {\em Phys. Rev.\/} {\bf 154} 138--149

\bibitem{Borghesani1990}
Borghesani A~F and Santini M 1990 {\em Phys. Rev. A\/} {\bf 42} 7377--7388 

\bibitem{Allen1982}
Allen M and Raabe O 1982 {\em J. Aerosol Sci.\/} {\bf 13} 537--547

\bibitem{Priest1981}
Priest G and Read S 1981 {\em Mind\/} {\bf 90} 274--279

\bibitem{katti}
Katti R, Jacobsen R, Stewart R and Jahangiri M 1986 {\em Adv. Cryo. Eng.\/}
  {\bf 31} 1189--1197

\bibitem{Knudsen1911}
Knudsen M and Weber S 1911 {\em Ann. Phys.\/} {\bf 341} 981--994 

\bibitem{Kim2005}
Kim J, Mulholland G, Kukuck S and Pui D 2005 {\em J. Res. Natl. Inst. Stand.
  Technol.\/} {\bf 110} 31--54

\bibitem{Reif}
Reif F 1985 {\em {Fundamentals of Statistical and Thermal Physics}\/} 
  (Auckland: McGraw-Hill)

\end{thebibliography}
\end{document}